# Cosmic Ray Induced Mass-Independent Oxygen Isotope Exchange: A Novel Mechanism for Producing $^{16}$O depletions in the Early Solar System

G. Dominguez[1*], J. Lucas[1], L. Tafla[1], M.C. Liu[2], K. McKeegan[2]

[1]California State University, Department of Physics, San Marcos, CA, 92096-0001

[2]Department of Earth, Planetary and Space Sciences, University of California–Los Angeles (UCLA), Los Angeles, CA 90095–1567, USA.

*Correspondence to: gdominguez@csusm.edu

**A fundamental puzzle of our solar system's formation is understanding why the terrestrial bodies including the planets, comets, and asteroids are depleted in $^{16}$O compared to the Sun. The most favored mechanism, the selective photodissociation of CO gas to produce $^{16}$O depleted water, requires finely tuned mixing timescales to transport $^{16}$O depleted water from the cold outer solar system to exchange isotopically with dust grains to produce the $^{16}$O depleted planetary bodies observed today. Here we show that energetic particle irradiation of $SiO_2$ (and $Al_2O_3$) makes them susceptible to anomalous isotope exchange with $H_2O$ ice at temperatures as low as 10 K. The observed magnitude of the anomalous isotope exchange ($\Delta^{17}$O) is sufficient to generate the $^{16}$O depletion characteristic of the terrestrial bodies in the solar system. We calculated the cosmic-ray exposure times needed to produce the observed $^{16}$O depletions in silicate ($SiO_2$) dust in the interstellar medium and early solar system and find that radiation damage induced oxygen isotope exchange could have rapidly (~$10^1$-$10^2$ yrs) depleted dust grains of $^{16}$O during the Sun's T-Tauri phase. Our model explains why the oldest and most refractory minerals found in the solar system, the anhydrous Calcium with Aluminum Inclusions (CAIs), are generally $^{16}$O enriched compared to chondrules and the bulk terrestrial solids and provides a mechanism for producing $^{16}$O depleted grains very early in the solar system's history. Our findings have broad implications for the distribution of oxygen isotopes in the solar system, the interstellar medium, the formation of the planets and its building blocks as well as the nature of mass-independent isotope effects.**

## Main Text:

The standard model of planetary system formation, informed by astronomical observations, suggests that protostars and planetary systems form when cold gas and dust in dense regions of molecular clouds collapse gravitationally. This infalling gas and dust feed the growth of the protostar until nuclear ignition commences and the surrounding gas and dust is pushed out by radiative pressure. Before this clearing stage is complete, the population of sub-micron interstellar dust grains and re-condensed solids derived from these grains must somehow collide, stick, and grow to become planetesimals (r~10-$10^2$ m) and ultimately the solid cores of gas-giants and terrestrial planets [1].

While astronomical observations provide key insights into the general process of planetary formation, the tightest constraints on the solar system's formation history are derived from the analysis of matter found within the solar system, including isotopes of major planet forming elements.

The discovery by NASA's Genesis mission that the solar system's terrestrial solids (e.g. the planets, asteroids, comets, and the bulk composition of meteorites) are depleted in $^{16}O$ by ~5-6% compared to the Sun was unexpected for several reasons[2]. First, interstellar silicate ($SiO_2$) dust grains originate from a variety of stellar outflows, each with distinct oxygen isotopic compositions. While residing in the interstellar medium (ISM), the isotopic signatures of individual dust grains, acquired from their stellar sources, are expected to be erased by destructive processes such as galactic cosmic ray irradiation and supernova shock wave sputtering and re-condensation before they are incorporated and preserved in planetary materials[3]. The homogeneous isotopic composition of minor elements in planetary materials, whose variations range from undetectable to <0.1%, strongly suggests that presolar dust grains were largely well-mixed before being incorporated into larger solid bodies in the SS[4]. Second, when the ratios of $^{17}O/^{16}O$ and $^{18}O/^{16}O$ of major solar system oxygen reservoirs including the Sun are plotted on a triple-isotope plot (See Figure 1), the distribution of oxygen isotopes produce a slope ~1 pattern that cannot be explained by conventional mass-dependent physical-chemical processes such as diffusion or thermally-driven isotope exchange[5]. Understanding when, where, and how the building blocks of the planets acquired their systematic $^{16}O$ depletion (or equivalently $^{17}O$ and $^{18}O$ enrichments) compared to the Sun is a major unsolved problem in planetary formation.

Here, we report on the discovery of an efficient and novel isotope exchange mechanism that by itself can explain how the solar system's solids acquired their $^{16}O$ depletion prior to the formation of the planets. Extrapolations of this mechanism to an early solar system with a T-Tauri-like Sun suggests that this process could have produced the $^{16}O$ depleted solids of the solar system in as little as $10^1$-$10^2$ years.

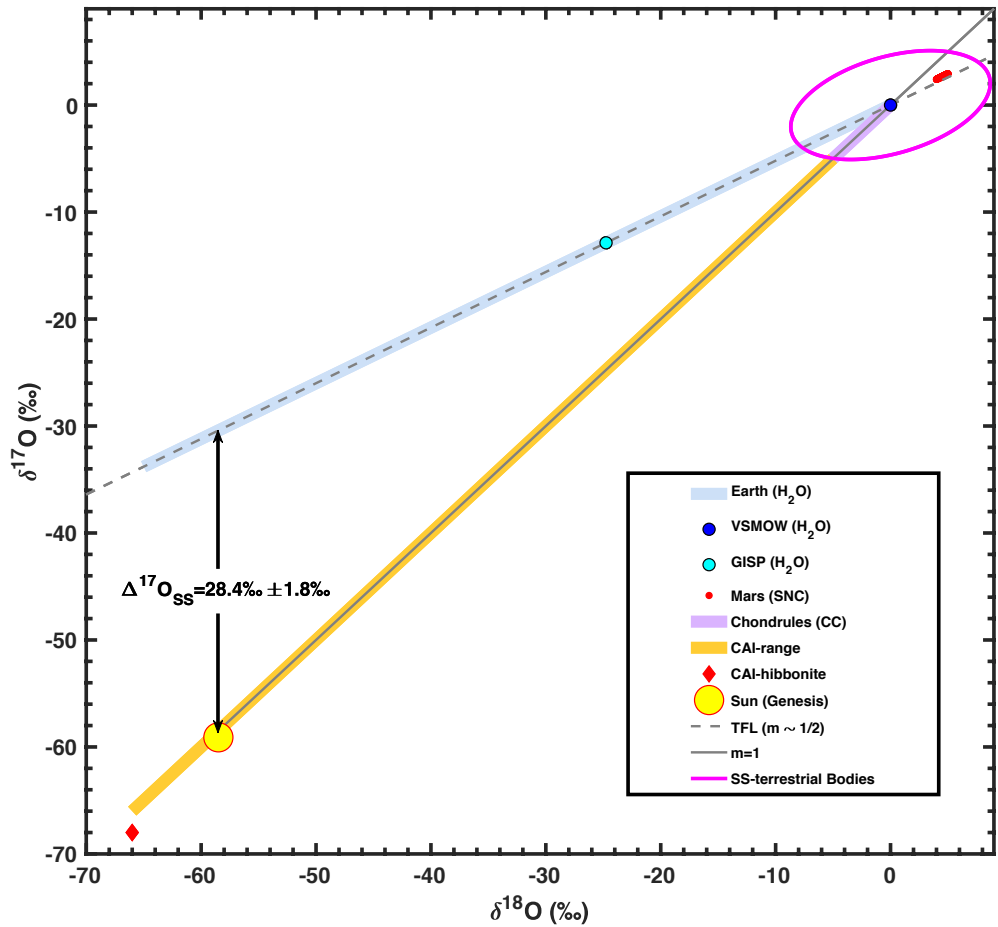

**Fig.1.** First order distribution of oxygen isotopes in the solar system, including major terrestrial $H_2O$ standards, such as Vienna Standard Mean Ocean Water (VSMOW) and Greenland Ice Sheet Standard Precipitation (GISP) in delta notation. These terrestrial water standards and non-atmospheric oxygen bearing terrestrial materials help define the Terrestrial Fractionation Line (TFL) for water and is produced by mass-dependent fractionation (MDF) processes. The bulk composition of Mars as inferred from laboratory measurements of SNC meteorites from Mars is slightly depleted in $^{16}O$ compared to Earth's bulk composition [6]. The Sun's composition as inferred by NASA Genesis[7]. Ranges of Allende chondrule and CAI compositions also shown as described by [8] and references therein. CAI-hibonite composition is from Isheyevo meteorite [9]. A key signature associated with MDF is that shifts in the relative abundances of $^{17}O/^{16}O$ and $^{18}O/^{16}O$, when quantified with respect to a standard (i.e.VSMOW) using delta notation as $\delta^{17}O$ and $\delta^{18}O$ respectively[i], a plot of $\delta^{17}O$ vs. $\delta^{18}O$ defines a line where $\delta^{17}O \sim \frac{1}{2}\ \delta^{18}O$. This contrasts with the "slope 1" line (m=1) that connects the bulk composition of the solar system, the Sun, with the terrestrial planets. Processes that selectively inject (or remove) $^{16}O$, produce slope 1 fractionation patterns that are traditionally quantified as $^{17}O$ anomalies, $\Delta^{17}O = \delta^{17}O$ -0.52x$\delta^{18}O$. In this notation, the Sun's isotopic composition, normalized to the Earth's ocean water (VSMOW), was inferred by the Genesis mission to be approximately given by $\delta^{18}O \sim -58.5‰$, $\delta^{17}O \sim -59.1$ ‰ and $\Delta^{17}O = -28.4‰+/- 1.8‰$. Conversely, when normalized to the Sun's composition, Earth's bulk composition (and those of other major solid bodies) are depleted in $^{16}O$ and characterized by oxygen isotopic compositions ($\Delta^{17}O$) ranging from $\sim 25$-32 ‰.

Models that attempt to explain the anomalous distribution of oxygen isotopes in the SS rely on identifying physical-chemical processes that selective produce or remove $^{16}O$ atoms and generally fall into two categories. The most popular, self-shielding of CO, postulates that the slope 1 pattern was produced by the selective photodissociation of $^{12}C^{17}O$ and $^{12}C^{18}O$ in select regions of the protoplanetary disk [10,11] or the molecular cloud from which the SS formed [8,11]. In essence, self-shielding mechanisms, by attenuating UV light across optically thick columns of CO, provide nearly pure sources of $^{17}O$ and $^{18}O$ atoms in select regions of CO clouds. Because the most abundant isotopologue ($^{12}C^{16}O$) is expected to be shielded from photodissociation, $^{17}O$ and $^{18}O$ produced from unshielded and less abundant $^{12}C^{17}O$ and $^{12}C^{18}O$ molecules provide a potential source of $^{16}O$ depleted water (ice) in astrophysical environments. This $^{16}O$ depleted water is then assumed, over some narrow range of timescales, to be transported to the warmer inner solar system where it can pass on its $^{16}O$ depletions to the building blocks of the planets (chondrules) via thermally-driven isotope exchange [12].

The second category, mass-independent chemistry, was proposed as a potential source of the slope 1 pattern when it was discovered that the formation of ozone ($O_3$) in the gas-phase, via $O_2 + O + M \rightarrow O_3 + M$, produces a slope 1 pattern [13]. While studies of gas-phase oxidation of SiO have recently been found to produce modest mass-independently fractionated (slope 1) reservoirs, the magnitude of the $^{17}O$ and $^{18}O$ enrichments or depletions, quantified as $\Delta^{17}O$, are insufficient to produce the SS's oxygen isotopic distribution (See Fig. 1) [14].

While the mass-dependent (m~1/2, See Fig. 1) nature of thermally-driven isotope exchange is well established [15], to the best of our knowledge, work focused on isotope exchange between volatile oxygen bearing ices like water and refractory solids like $SiO_2$ in astrophysical conditions has not been previously reported. Thermally-driven isotope exchange between water ($H_2^{16}O$, $H_2^{17}O$, $H_2^{18}O$) and dust grain silicates ($SiO_2$) in cold molecular cloud conditions is not expected to be significant since isotope exchange requires breaking Si-O, which have bond energies that are significantly larger (~10 eV or 120,000K) than the available thermal energy (10-80K). Recent work, however, has shown that cosmic-rays are capable of triggering non-equilibrium chemical reactions in molecular cloud core conditions [16]. In short, the energy required to overcome chemical activation energy barriers in cold (T~10K) astrophysical environments may be provided by the energy-loss of relativistic cosmic-ray ions traversing dust grains and producing a cascade of secondary electrons with more modest energies. This type of energy deposition, could in principle, also break Si-O bonds and, in the presence of other oxygen bearing volatile reservoirs, make them susceptible to oxygen isotope exchange.

To better understand non-thermal isotope exchange in astrophysical conditions, we conducted a series of experiments using isotopically tagged water-ice covered $SiO_2$ and $Al_2O_3$ solids irradiated with 5 keV electrons. We found that radiation damage of solid surfaces in the presence of a water-ice reservoir, even at 10 K, produces mass-independent oxygen isotopic isotope exchange with a magnitude that may explain the distribution of oxygen isotopes found in the solar system[17]. By considering the cosmic-ray intensities in a variety of astrophysical environments, we identified the early solar system's T-Tauri phase as a setting where cosmic-ray induced oxygen isotope exchange was most likely to have occurred. Our results have broad implications in planetary science and cosmochemistry, astronomy, and the physical chemistry of mass-independent isotope effects in nature.

## Methods

Electron induced isotope exchange at T=10 K between $H_2O$-ice and oxygen bearing solid surfaces were carried out at CSU San Marcos. Electron irradiation times of ~45 minutes (low e- fluence) and ~ 450 minutes (high e- fluence) were done using an electron gun (Kimball Physics) beam current of ~50 µAmps and energies of 5 keV/electron. We evaluated and quantified the amount of oxygen isotope exchange in the solids by using Secondary Ion Mass Spectrometry (SIMS) at UCLA. We focused on quantifying shifts in the isotopic composition of irradiated solid surfaces since shifts in the isotopic composition of the water-ice reservoir ($N_{H2O} > 3x10^{18}$) were expected to be small compared to those of metal oxide solid targets ($N_{M-O} \sim 2x10^{13}$) affected by the irradiation. We conducted irradiations with relatively low and high fluences (~45 vs ~450 min.) using a variety of targets and water compositions. Additional details, including instrumental blanks, are found in **Supplementary Information.**

## Results

SIMS analysis of both low and high fluence targets revealed measurable (and anomalous) shifts in their oxygen isotopic compositions. Figure 2 plots the isotopic compositions ($\delta^{17}O$ and $\delta^{18}O$) of ~ 50 x 50 micron raster squares on the surfaces of $SiO_2$ and $Al_2O_3$ targets that received a relatively low dose of 5 keV electrons ($\Phi \sim 2x10^{17}$ cm$^{-2}$). One of these samples ($SiO_2$-2) was first irradiated with 5 keV electrons and then exposed to $H_2O$ vapor with GISP isotopic composition while being held at T=10K. We find that the isotopic exchange produced isotopic fractionation patterns consistent with a slope 1 line ($\Delta\delta^{17}O/\Delta\delta^{18}O \sim 1$).

Figure 3 plots the results of SIMS measurements of the spots on a quartz sample ($SiO_2$-9) that was covered in water-ice (GISP) first and then exposed to a 5 keV electron beam for ~10 times longer ($\Phi \sim 2.5x10^{18}$ cm$^{-2}$) followed by five additional exposures to water-vapor (GISP) while being held at T=10K. Here we find that large $^{16}O$ depletions that are ~10 times the size of those seen in the low-fluence experiments. Additional details on SIMS analysis and water standards are provided in **Supplementary Information.**

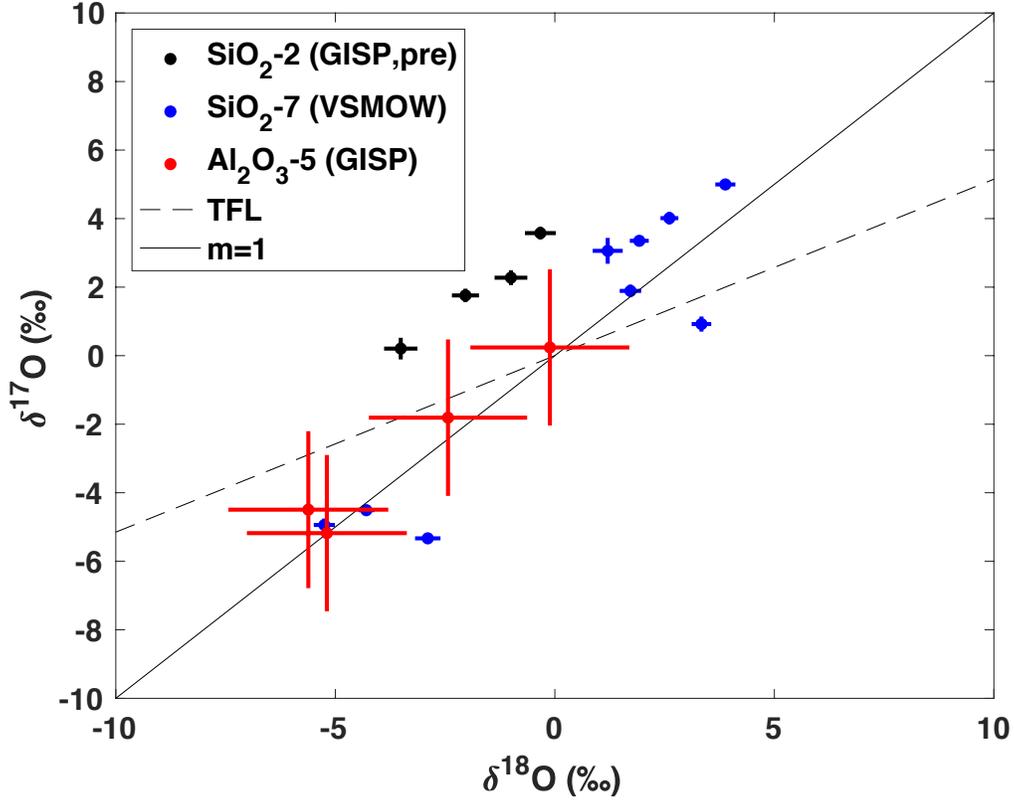

**Fig.2.** $\delta^{17}O$ vs. $\delta^{18}O$ of individual analyzed craters of irradiated surfaces as measured by SIMS and corresponding 2σ uncertainties. Composition is reported relative to the corresponding un-exposed surfaces. Lines with slopes equal to the terrestrial fractionation line slope (TFL-slope, $\Delta\delta^{17}O/\Delta\delta^{18}O \simeq 0.515$) and a $\Delta\delta^{17}O/\Delta\delta^{18}O = 1$ trend are shown for reference. Determinations of $\Delta^{17}O$ in $Al_2O_3$ and quartz have estimated uncertainties of $\simeq \pm 0.65$ ‰ and $\simeq \pm 0.35$ ‰ . These uncertainties were determined by adding the standard error of the cycle-to-cycle variation in δ values and their spot-to-spot variation standard error obtained on blank samples in quadrature. Best fit to overall data yields a mass-independent slope of 0.98 +/- 0.27 (2 σ). See **Supplementary Information** for isotopic composition of an unexposed $SiO_2$ blank.

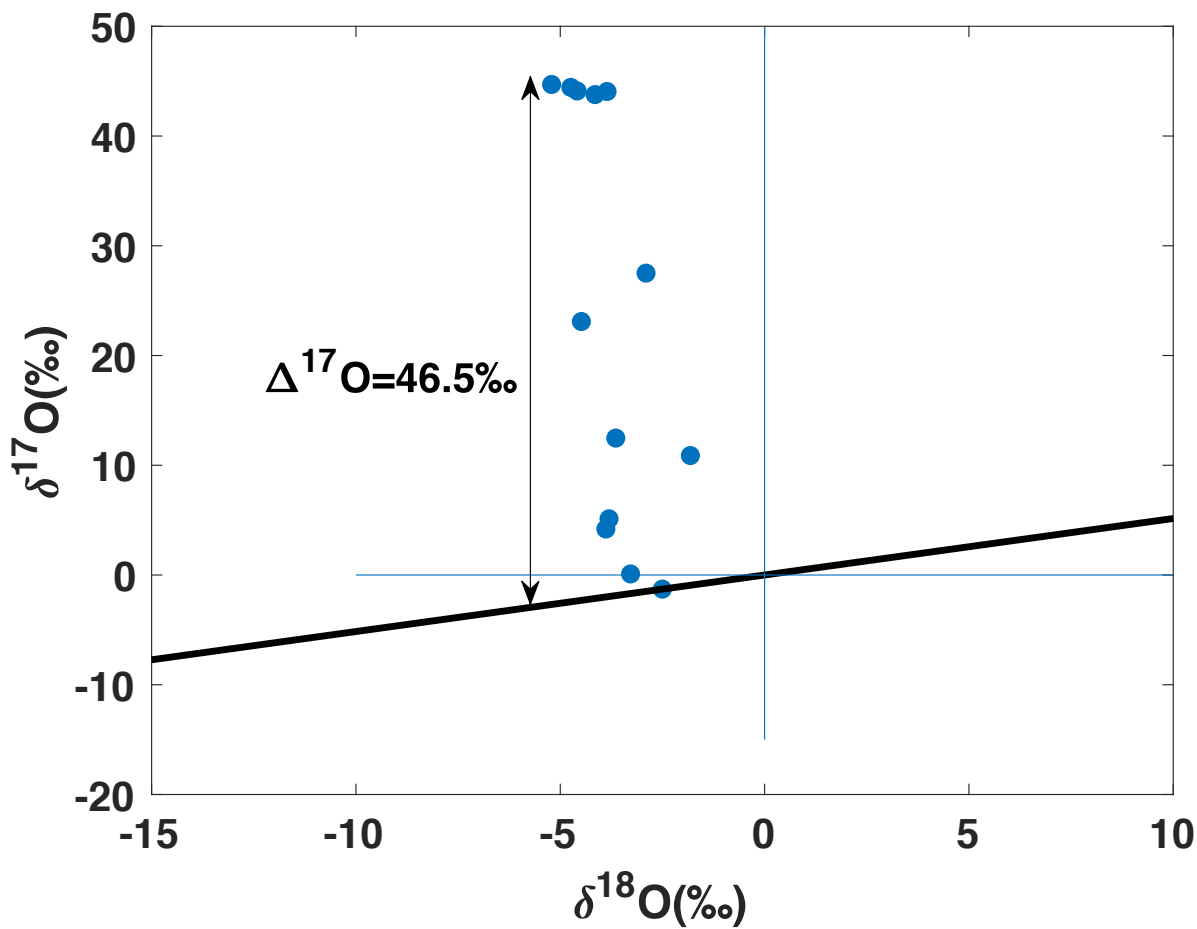

**Fig.3.** Triple-isotope plot of a $SiO_2$-quartz sample exposed to ~450 minutes of ~50 μA e- beam at T=10 K. We note the significant mass-independent fractionation ($\Delta^{17}O$) of the surface. We interpret the non-slope 1 as an artifact of secondary mass-dependent fractionation of the target or secondary oxygen bearing compounds that were formed. This interpretation is consistent with the defect formation calculations (See **Supplementary Information**) which suggest ~ 100% of the Si-O bonds near the surface were disrupted by the e-beam during this high-fluence experiment. Solid black line represents mass dependent fractionation pattern for data that are normalized with respect to sample blank. $\Delta^{17}O$=46.5‰ +/- 0.6‰ is the average and standard deviation of the most $^{16}O$ depleted spots.

## Discussion

Our results presented in Figure 2 demonstrate that ion irradiation of solids at low temperatures produce conditions that result in oxygen isotope exchange between $H_2O$-ice and the underlying solid. This isotope exchange produces a slope ~1 (m=0.98 +/- 0.27 (2σ)) pattern that resembles the oxygen isotopic distribution of the solar system summarized in Figure 1 and suggests that molecular compounds with $^{16}O$ depletions and enrichments are produced as a result of the low-temperature irradiation of these surfaces. To our knowledge, this is the first time that direct evidence of non-equilibrium, mass-independent chemistry has been found in low-temperature ion irradiation experiments. Furthermore, the observation that water need not be simultaneously present with the irradiation strongly suggests that defect formation resulting from the energy-loss of ions, regardless of whether the source is a proton or $H^+$ or $e^-$, is the essential factor necessary for producing the anomalous oxygen isotope exchange patterns we report.

While the energy of electrons used in our experiments (5 keV) are not as high as the energies of galactic or solar cosmic rays (E>10 MeV), we believe that our experiments adequately simulate the essential features of cosmic-ray induced damage in solids. First, relativistic protons moving through solids lose energy primarily via Coulomb interactions with electrons bound to atoms in the lattice[18]. For a typical cosmic-ray and interstellar dust grains ($r_g$~200 nm), this energy loss is consistent with the energy of electrons we used (~$10^2$-$10^3$ eV). CR particle ($e^-$, $H^+$, $He^{2+}$) interactions with solids produce a cascade of secondary electrons with lower but sufficient energies (~10-$10^4$ eV) to disrupt chemical bonds, producing defects. This cascade of secondary electron production is expected to continue until an energetic threshold ($E^*$) is reached and the secondary electrons no longer have the energy required to dislocate atoms or other electrons. Thus, while our experimental work makes use of electrons, we believe that future work with protons as the source of secondary electrons are likely to yield similar results to the ones we report here.

Our finding that the magnitude of $^{16}O$ depletion ($\Delta^{17}O$) of the solid scales with electron fluence (or energy) received by the sample in Figure 3 is consistent with the theoretical expectation that the amount of radiation damage, or atomic dislocations, produced in the solid target is a function of the *total amount of energy deposited*, as has been previously observed for Si [19].

The non-uniform $\delta^{18}O$ and $\delta^{17}O$ isotopic compositions of the surfaces in Figure 2 may result from several factors. First, it is possible that the water-ice thickness variations could, in principle, lead to differences in the amount of electron fluences that penetrated the surface of the samples that were covered in water-ice prior to irradiation with the electron beam. The non-uniform isotopic compositions observed in the pre-irradiated surface, however, argues against this explanation. We suggest, instead, that the isotopic variations were more likely produced by the presence of two (or more) volatile oxygen reservoirs that were produced on the irradiated surfaces. Molecular Dynamics (MD) simulations of radiation damage applied to our experiments suggest that only a small fraction (fD) of the Si-O bonds in the irradiated volume were broken by the electron beam (See Supplementary Information). These defects would be susceptible to being re-oxidized by any oxygen bearing volatile compounds found on the surface. Using mass balance and the assumption that there were two co-existing $^{16}O$ depleted reservoirs of equal magnitude (+$\Delta^{17}O_{MIF}$, -$\Delta^{17}O_{MIF}$), we can write the oxygen isotopic composition of a SIMS analytical volume as:

$$< \Delta^{17}O_{SIMS} >_V = f_+ \cdot \Delta^{17}O_{MIF} + f_- \cdot (-\Delta^{17}O_{MIF}) + (1 - f_D) \cdot \Delta^{17}O_{blank} \qquad (1)$$

where $f_+$ (=$f_D$- $f_-$) and $f_-$ are the fraction of broken Si-O bonds ($f_D$) in the SIMS analytical volume that were re-oxidized by the +$\Delta^{17}O_{MIF}$ and -$\Delta^{17}O_{MIF}$ reservoirs respectively and $\Delta^{17}O_{blank}$ is the anomaly of the unaltered sample ($\Delta^{17}O_{blank}$ = 0). We used this interpretation to obtain a *lower limit* on the instantaneous fractionation factor ($\Delta^{17}O_{MIF}$) associated with the isotope exchange by taking the value of the most isotopically (positive) anomalous spot of a sample in the low fluence experiments, set f-=0, and solved the above equation for +$\Delta^{17}O_{MIF}$ to get:

$$\Delta^{17}O_{MIF} \geq \frac{\max(<\Delta^{17}O_{MIF}>_V)}{f_D} \quad (2)$$

$$\Delta^{17}O_{MIF} \geq \frac{(3.5\pm1)}{f_D}\text{‰} \quad (3)$$

We note that the equality corresponds to the case where $f_D$=1. Estimates of $f_D$ using a Molecular Dynamics simulations of electron energy deposition and defect production in quartz targets (**See Supplementary Information**) [20] suggests that $f_D$ may be as low as 0.10 in the low fluence experiments, which implies that $\Delta^{17}O_{MIF} > 35$‰. Additional experiments are needed to better constrain $f_D$ (in progress).

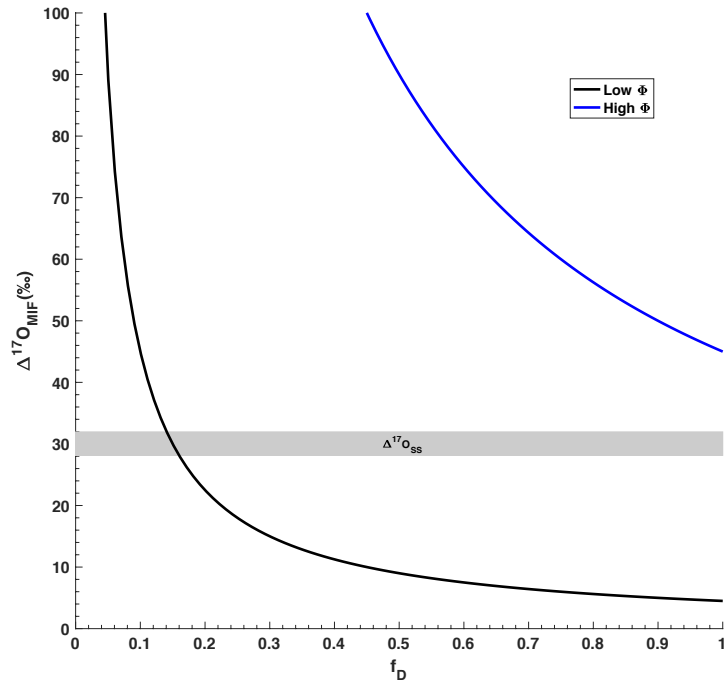

**Fig.3.** Analysis of instantaneous fractionation factor ($\Delta^{17}O_{MIF}$) inferred from the SIMS measurements of the low and high fluence quartz surfaces as a function of the fraction of Si-O bonds ($f_D$) that have exchanged isotopically. Shaded gray region denotes the magnitude of the anomaly that separates the Sun from the terrestrial bodies of the solar system [7] (See Figure 1).

The larger maximum (and range) of isotopic anomalies observed in the high fluence experiment is consistent with the expectation that increasing an electron dosage (energy deposition) would lead to more broken Si-O bonds breakage and overall oxygen isotopic exchange. Here, we found that an increase in the exposure duration of irradiation by 5 keV electrons by a factor of ~10 increased the amount of $^{16}O$ depletion recorded in the solid by a similar factor, as evidenced by several SIMS spots with measured isotopic anomalies ($\Delta^{17}O_{SIMS}$) as large as 47.3‰. To our knowledge, these spots represent the largest oxygen isotopic anomalies transferred to silicates observed in the laboratory and imply that the instantaneous fractionation factor associated with the mass-independent chemistry on the cold surfaces had $\Delta^{17}O_{MIF} \geq 47.3‰$ We plot $\Delta^{17}O_{MIF}$ as a function of $f_D$ and compare these values to the $^{16}O$ depletion that characterizes the bulk of the solar system's terrestrial solids ($\Delta^{17}O_{SS}$) as reported by the Genesis mission in Fig.3. We conclude that we have identified a novel physical-chemical process that can occur at temperatures as low as 10 K and that by itself may produce solids with the systematic $^{16}O$ depletion observed in the solar system. We summarize the oxygen isotopic anomalies observed our experiments as a function of electron fluence in Figure 4 below.

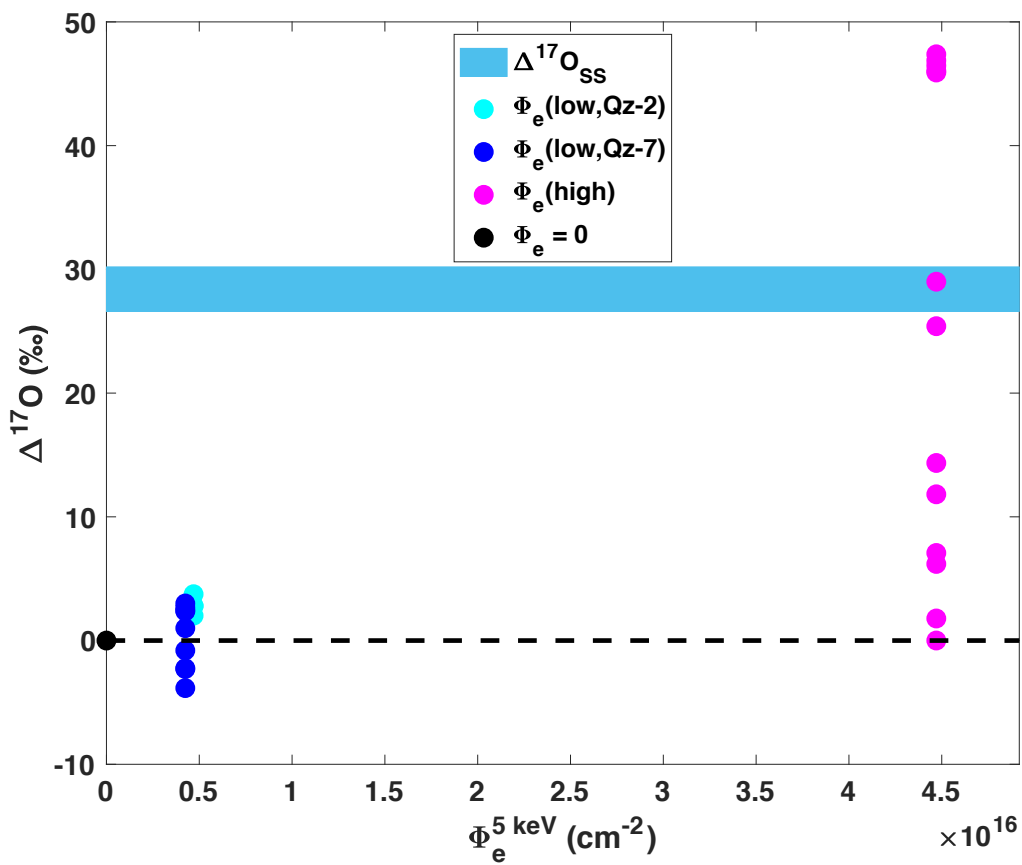

Fig.4. Summary of isotopic anomalies ($\Delta^{17}O$) as a function of 5 keV electron fluences. $\Delta^{17}O$ normalized with respect to unexposed blank materials (See Fig. S2 for oxygen isotopic analysis of blank $SiO_2$).

To evaluate whether the isotope exchange process we have discovered is a plausible mechanism for producing the $^{16}O$ depleted solids found in the solar system, we estimated the timescales necessary for dust grains to experience comparable energetic particle exposures in the ISM and in the early solar system. A detailed summary of the CR intensity energy spectra, energetic fluxes, and integration methods used for our estimates are included in **Supplementary Information** (see Figure S5) and we briefly summarize these here. We estimated CR proton fluxes in the ISM (5-10 cm$^{-2}$ s$^{-1}$) and the solar system (~150 cm$^{-2}$ s$^{-1}$) using observations made by the Voyager spacecraft[21] and observations of Solar Energetic particle (SEP) events associated with Coronal Mass

Ejections (See **Supplementary Information**) respectively[22]. Observations of T-Tauri stars, self-shielding models [10], and measurements of radionuclides in meteorites [23,24] suggest that the energetic output of the Sun may have been a factor of $\sim10^4$-$10^5$ higher in the early solar system [25,26]. Therefore, to assess the exposures experienced by dust grains during the Sun's T-Tauri phase, we scaled contemporary solar CR output estimates by a factor of $\sim10^5$ [25]. Using these fluxes we estimate that exposure times of $\sim10^{17}$ s ( $>10^3$ Myrs) in the ISM, $\sim10^{16}$ s ($\sim$424 Myrs) in the contemporary solar system, and $\sim10^{11}$ s ($\sim10^3$ - $10^4$ yrs) in the early solar system would be required to expose dust grains to energetic particle fluence ($\sim$2x$10^{18}$ cm$^{-2}$ ) that characterize our high fluence experiments.

We performed more detailed estimates of exposure times as follows. Galactic and solar cosmic ray proton and electron energies, on average, exceed 10 MeV. At these energies, the stopping length ($>10^{-2}$ m) of typical CRs greatly exceeds the typical size of interstellar dust grains ($r_g\sim10^{-7}$ m) [18]. Thus, the relevant energy for evaluating how much radiation damage is experienced by dust grains is the energy deposited per cosmic-ray ( $E_D \sim \dfrac{dE}{dx} 2r_g$ ) integrated over the spectral intensity ( $\dfrac{dJ}{dE}$ ) of cosmic rays. For example, the dE/dx of a 40 MeV proton in SiO$_2$ is $\sim$39.23 MeV/cm, which translates into an energetic deposition of $\sim$790 eV in a 200 nm dust grain per cosmic-ray proton interaction[18]. Adopting a simple energy threshold model that requires 100 eV of energy deposition per defect, we estimate that a single CR proton can produce between 8-9 atomic dislocations ($N_D$) in a dust grains with $r_g\sim$200 nm. A more detailed analysis that considers the spectral intensity of cosmic ray protons together with the CR energy dependence of energy loss and defect formation in SiO$_2$ is detailed in **Supplementary Information**. In this more careful analysis, the rate of defect formation a dust grain of radius $r_g$ is given by:

$$\frac{dN_D(r_g)}{dt} = 4\pi \int_{E_{min}}^{E_{max}} N_D(E_{CR}, r_g) \frac{dJ(E_{CR})}{dE} r_g^2 dE_{CR} \qquad (4)$$

where $N_D$ ($E_{CR}$) is the number of defects produced by a cosmic ray with energy $E_{CR}$ traversing a silicate dust grain with radius $r_g$. The (1/e) timescale for resetting the Si-O bonds of a water-ice covered SiO$_2$ dust is given by:

$$\tau_D = \frac{N_O}{\left(\dfrac{dN_D}{dt}\right)_{Total}} \qquad (5)$$

where $N_O$ is the number of oxygen atoms in a dust grain. Our analysis finds that the timescale for producing $^{16}$O depleted solids early in the solar system may have been as rapid as $10^1$-$10^2$ years in regions of the early solar system where water ice covered dust grains ($r_g<$200 nm) would be expected to exist, beyond the snowline (R>3AU, See Figure 5). In contrast, isotopic exchange timescales in the ISM are $\sim 10^5$ times slower (1-10 Myrs), suggesting a minor but potentially observable role for CR induced isotope exchange in molecular clouds over their lifetimes ($\sim$30 Myrs).

We now consider the consequences of having rapid isotope exchange timescales in the early solar system. In a closed system where radiation induced isotope exchange occurs relatively quickly (i.e. mixing timescales>> $\tau_D$ ), mass-balance between the volatile oxygen reservoir (O-ice) and solid dust grains (s) requires that (See **Supplementary Information** ):

$$f_s \Delta^{17}O_s + f_{ice}\Delta^{17}O_{ice} = 0 \qquad (6)$$

$$\Delta^{17}O_s - \Delta^{17}O_{ice} = \Delta^{17}O_{max} \qquad (7)$$

where the $^{16}O$ excess for the entire system (volatiles + dust) is set to zero and $f_s$ and $f_{ice}$ are the fraction of all oxygen atoms found in the dust and water ice phases respectively (See **Supplementary Information**). Equations (6) and (7) imply that:

$$\Delta^{17}O_s = f_{ice}\Delta^{17}O_{max} \qquad (8)$$

To illustrate, if we set $\Delta^{17}O_{max} \simeq 47.3$ ‰, the largest $^{16}O$ depletion detected in our high fluence experiment, and we set $f_s{\sim}0.25$ and $f_{ice}{\sim}0.75$ as suggested by the cosmochemical abundances of Si and O and the average elemental composition of interstellar dust ($Mg_{1.32}Fe_{0.10}Na_{0.01}SiO_{3.45}$) [27,28], the equilibrium oxygen isotopic composition of the irradiation exchanged dust would be $\Delta^{17}O_s \simeq 35.4$ ‰ compared to the isotopic composition of bulk oxygen isotopic composition of the SS (Sun). Interestingly this $^{16}O$ depletion spans the entire range of $^{16}O$ depletions observed in major meteorite classes [29]. We note, however, that if 20% of the oxygen atoms are found in a reservoir that does not participate in isotope exchange with the water-ice or silicate grains (e.g. CO gas), then $f_{ice}$=0.60 and $\Delta^{17}O_s \simeq 28$ ‰, which is consistent with $\Delta^{17}O_{ss}$ found by Genesis $\Delta^{17}O_{ss} = 28.4‰ \pm 1.8‰$ ) for the terrestrial bodies. Sequestration of interstellar dust grains within larger planetesimals before isotopic equilibrium is reached would produce more modest $^{16}O$ depletions ($\Delta^{17}O{<}35.4$ ‰). Future work will need to explore how transport, chemical evolution of the PPD, self-shielding of CO, and dust grain sequestration contribute to the range of $^{16}O$ depletions of terrestrial bodies forming in different SS environments (See Figure 1).

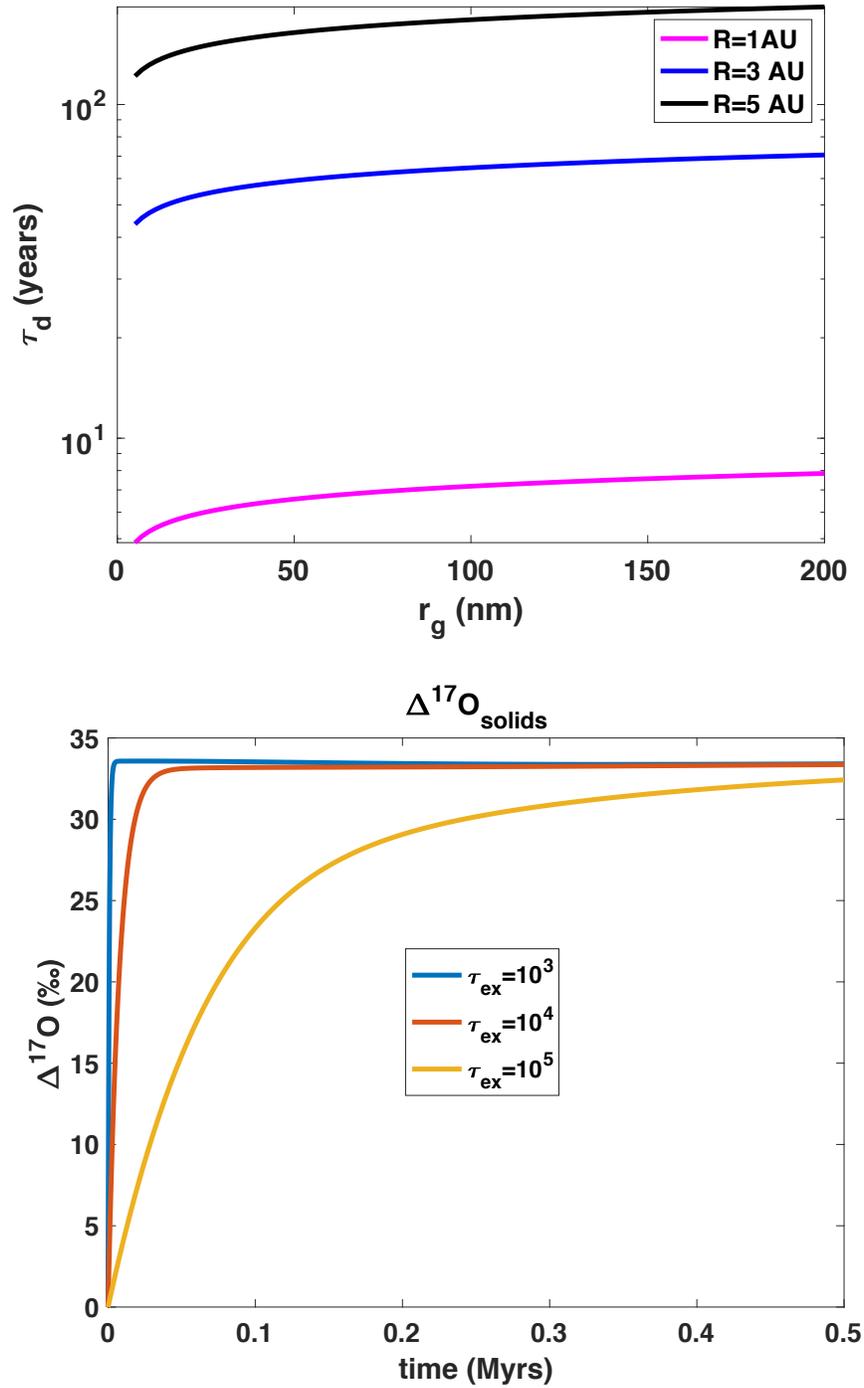

**Fig.5.** (top) Defect formation timescales ($\tau_D$, Equation (5)) for dust grains exposed to a T-Tauri cosmic-ray proton intensities as a function of dust grain size and distance from the Sun (1 AU=1.5x10$^{11}$ m). (bottom) Time evolution of the oxygen isotopic composition ($^{16}$O depletion) of water-ice covered dust grains vs. time due to radiation induced isotope exchange assuming instantaneous fractionation factor of $\Delta^{17}O_{max} \simeq 47.3$ ‰ and isotope exchange that is rate limited

by defect formation timescale ($\tau_{ex} = \tau_d$). Isotopic composition of dust and gas set equal to the bulk ($\Delta^{17}O = 0\text{‰}$) at t=0 but final results are insensitive to assumptions about the initial composition of the dust relative to the bulk.

Our model for producing the $^{16}O$ depletion of solar system's terrestrial bodies has several attractive features compared to self-shielding models. First, a slope 1 single step fractionation mechanism provides a way for producing terrestrial bodies with relatively uniform oxygen isotopic compositions without the need for finely tuned mixing timescales or selective transport of $H_2O$ from the outer regions of the solar system[10]. Second, our model explains why the oldest minerals found in the solar system that formed in the hot inner solar system, *anhydrous* CAIs, are $^{16}O$ enriched and closely resemble the Sun's oxygen isotopic composition. In contrast, our model suggests that interstellar dust grains that eventually become melted to form chondrules, the building blocks of the planets, first underwent isotope exchange beyond the SS's snow-line (>3 AU) and became $^{16}O$ depleted there. Finally, our model can produce $^{16}O$ depleted dust at the earliest stages of protoplanetary disk formation in as little as $10^1$-$10^2$ years [30].

As shown by our high fluence experiments, $^{16}O$ depleted solids ($\Delta^{17}O>0$) can be preferentially produced in cold astrophysical conditions in the presence of water ice and energetic particle irradiation. While understanding the preferential retainment of the $^{16}O$ depleted reservoir in these experiments is beyond the scope of this work, the observed isotopic separation is likely the result of differences in the residence time and chemical reactivity of the molecules that carry the $^{16}O$ enrichments and depletions. Additional experimental work will be needed to identify these compounds to provide additional constraints on the formation environments and history of meteorites and asteroids. Identification of the isotopically anomalous compounds will also be of interest to quantum symmetry-based physical-chemical models of $^{16}O$ enriched and depleted reservoirs on surfaces [31]. Independent of interpretation, our findings confirm the hypothesis that non-equilibrium chemistry of volatile compounds on cold dust grain surfaces exists and has sufficient magnitudes to potentially explain the $^{16}O$ depletion of the planets [17].

We have found that radiation damage of interstellar dust caused by cosmic ray energy deposition together with interactions with water ice can create anomalous, mass-independently fractionated oxygen reservoirs in astrophysical conditions. The magnitude of the isotopic anomaly is sufficient to produce the $^{16}O$ depletions of the terrestrial bodies found in the solar system. While our work here does not preclude a role for self-shielding in explaining some of the solar system's oxygen isotopic variations, it establishes that self-shielding is not the only way of producing *significant* oxygen isotopic anomalies in astrophysical environments [32]. The rapid-timescales for producing $^{16}O$ depleted and enriched reservoirs in the early solar system at the snowline and beyond deserve consideration by models seeking to explain the distribution of oxygen isotopes in the solar system or within individual CAIs that appear to have recorded short lived $\Delta^{17}O$ variations early in the solar system's formation history [33]. Our results have implications for the oxygen isotopic composition of water and solids found in the permanently shadowed regions of the moon where water-ice covered solids are expected to be exposed to solar cosmic-rays over several millions to billions of years [34].

## Supplementary Information

Electron induced isotope exchange at T=10 K between $H_2O$-ice and oxygen bearing solid surfaces was carried out using the Isotopic Characterization Experiment (ICE) at CSU San Marcos. Electron irradiation times of ~45 minutes (low e- fluence) and ~ 450 minutes (high e-fluence) were done using an electron gun (Kimball Physics) beam current of ~50 µAmps and energies of 5 keV/electron. These fluences and energies were chosen to maximize the number of secondary electrons, beam penetration, and defects created in the solid targets. We note that while the temperature of the sample holder (See Fig. S1) indicated that T=10K on the edge of the target (thermometry provide by a calibrated Si diode and Lakeshore LS-335 temperature controller), it is possible that the temperature at the center of the target may have systematically warmed up as a portion of the energy deposited by 5 keV electrons was converted to thermal energy via lattice vibrations.

Isotope ratio determinations of blank and electron irradiation exposed $Al_2O_3$ (sapphire) and $SiO_2$ (quartz) surfaces were performed using the CAMECA IMS-1290 secondary ion mass spectrometer at UCLA. In some experiments, water vapor was deposited and frozen out prior to electron irradiation and in others it was deposited after the electron irradiation was completed.

<u>Materials</u> <u>Used</u>
Electron irradiation targets, optical quality disks 10 mm diameter and 1-3 mm thick, made of either ($SiO_2$-quartz) or $Al_2O_3$ (sapphire) were acquired from ColdEdge Technologies to fit into the optical substrate holder (See Fig. S1).

**V**ienna **S**tandard **M**ean **O**ceanic **W**ater (VSMOW) and **G**reenland **I**ce **S**heet **P**recipitation (GISP) water standards were acquired from National Institutes of Standards and Technology (NIST). Their isotopic compositions are listed in Table S1.

| $H_2O$ Standard | $\delta^{18}O$ x1000 | $\delta^{17}O$ x1000 | $\delta^{2}H$ x1000 |
|---|---|---|---|
| **VSMOW** | 0 ‰ | = 0.52 x $\delta^{18}O$ | 0 ‰ |
| **GISP** | $-24.78 \pm 0.09$‰ | = 0.52 x $\delta^{18}O$ | -189.5 ‰ |

*Table S1. Water standards used in experimental work described below and in main text.*

### Simultaneous Exposure Experiments

At the start of a new target exposure experiment, the cryocooler's 6-inch CF flange is unbolted and the cold-head from the UHV chamber is carefully withdrawn (See Figure S1). New target disks are carefully[1] placed inside a cylindrical well in the gold-plated sample holder attached to the cold-head and secured using a gold-plated cover piece, indium wire, and UHV rated screws.

---

[1] Using UHV standard techniques including facemasks, beard covers (when applicable), hair nets and UHV rated gloves.

Once secured and inspected for fingerprints and other contamination, the cold-head and sample holder are remounted carefully and aligned to maximize electron beam fluence.

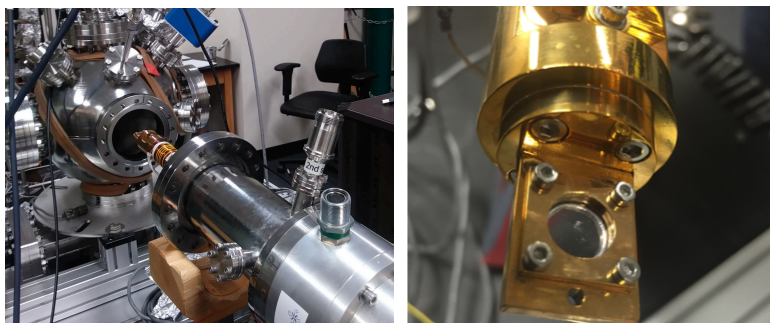

**Fig. S1.** (Left) View of UHV chamber and cryocooler with cold-head and sample holder. (Right) Close up of sample holder with $Al_2O_3$ window and mounting cover and Si diode thermistors for temperature monitoring.

The first set of experiments that we completed were simultaneous irradiation of $H_2O$-ice and solid surfaces. To produce these water-ice covered surfaces we first pumped out the heated (T=40 degrees C) UHV chamber overnight until the pressure was P~$10^{-9}$ Torr or less in order to minimize unwanted $H_2O$ backgrounds acquired from venting the chamber. We prepared water ice samples by injecting 5-10 µL of a GISP or VSMOW water standard into the vacuum line followed by controlled deposition of this water onto the cold solid surface using a stainless-steel needle valve.

Once the pressure inside the UHV stabilized below ~ $2 \times 10^{-8}$ Torr, we initiated electron irradiation and monitored the electron flux every 10 minutes for 30 seconds using a Faraday cup, pico-Ammeter, and MATLAB data acquisition software developed at CSUSM. This monitoring is necessary as adjustments sometimes need to be made to the cathode voltage in order to maintain a steady beam current. Once electron irradiation was complete, the He compressor was turned off and the cold surface was brought up to room temperature while simultaneously pumping the UHV chamber using the Shimadzu magnetically levitated Turbo pump (TMP-303LM). Because we wanted to minimize water-liquid-vapor interactions, we did not collect the residual $H_2O$ for the experiments reported here. Pressures less than ~ $3 \times 10^{-8}$ Torr were maintained throughout the electron bombardment, minimizing the potential for any gas-phase chemistry.

### Pre-exposure Experiment

One sample ($SiO_2$–2) was pre-irradiated and then exposed to a total of 30 µL of GISP ($\delta^{18}O = -24.78 \pm 0.09‰$) $H_2O$ water vapor while being held at 10 K. Once irradiations and water-ice exposures were done, samples were stored in clear plastic containers until they were analyzed at the NSF UCLA SIMS facility.

### High Fluence Experiment

The high fluence experiment ($SiO_2$-9) was conducted using similar protocols as the low-fluence experiment ($SiO_2$-7) using 5 microLiters of GISP water to coat the surface followed by ~450 minutes of e-beam exposure. After electron beam exposure, we did five rounds of 5 microLiters of GISP water vapor deposition while holding the surface at T = 10 K.

**SIMS Measurements (UCLA)**

The oxygen isotopic composition of irradiated and non-irradiated (standard) surfaces and near surface regions was determined using the UCLA IMS-1290 ion microprobe. The samples were raster-sputtered (50 µm x 50 µm) with a 3 nA $Cs^+$ primary ion beam by collaborator. Secondary ion intensities were measured in multi-collection mode under a mass resolution (M/$\Delta$M) of 6,000, with $^{16}O$, $^{17}O$ and $^{18}O$ being collected with L'2 (Faraday cup, $10^{10}$ $\Omega$), axial electron multiplier (EM) and H1 (Faraday cup, $10^{11}$ $\Omega$), respectively. A 35 µm by 35 µm field aperture was used to ensure that only signals from the center of a crater were collected. No pre-sputtering was applied prior to data acquisition. Non-irradiated quartz and sapphire were used as standards to infer the

**SIMS Standards (UCLA)**

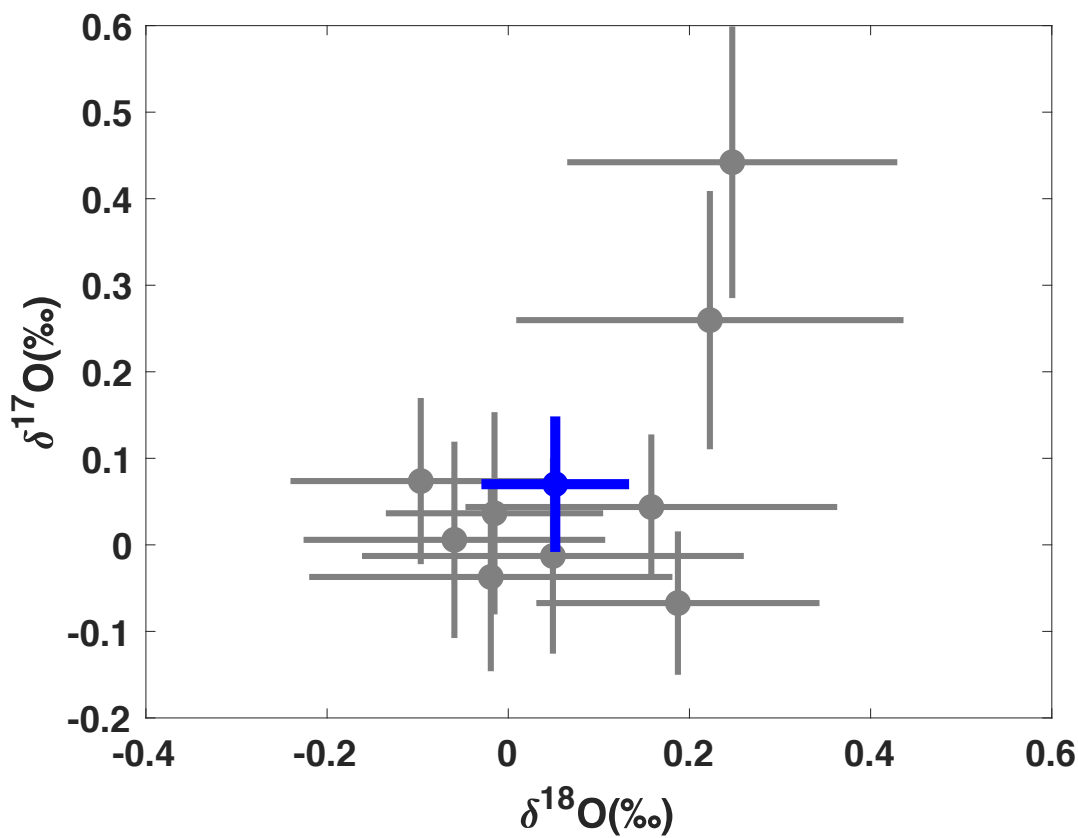

Fig. S2. SIMS analysis spot (gray)s of unirradiated quartz ($SiO_2$) measured in delta units with respect to the weighted average of these spots (shown in blue).

deviation of oxygen isotopic compositions from a mass-dependent fractionation line (See Fig. S2 for quartz blank).

**Radiation Defects in Quartz**

The creation of defects in solids like quartz has been studied experimentally and theoretically. While there remain uncertainties in this understanding, it appears that the net effect of energy losses by ions in solids is to permanently disrupt the structure of ordered materials like quartz, leading to the creation of several kinds of "defects". Wang et al., using Molecular Dynamics (MD) to simulate radiation damage caused by ballistic energy inputs into α-quartz, note that in addition to oxygen vacancies, which are associated with the displacement of oxygen atoms from their lattice positions, radiation damage also produces over-coordinated oxygen and five-coordinated Si atoms[20]. While it is unclear how the various types of defect types contribute to the isotopic signals we observe in our low and high fluence experiments, here we focus on evaluating the plausibility of creating atomic dislocations of the type that would be most susceptible to isotopic exchange with volatile oxygen.

First, using conservation of energy and momentum, we estimate the maximum energy of a head-on collision between an electron ($m_e$) and an $^{16}$O atom ($16m_p$), in what is referred to as a Primary Knock-on Atom (PKA) event. This maximum recoil energy, for an electron with E=5 keV colliding head on with an $^{16}$O atom, is given by:

$$E_R = \tfrac{1}{2}\left(16m_p\right)v_R^2 \tag{9}$$

where the recoil velocity ($v_R$) is given by:

$$v_R = \frac{(2m_e)}{(m_e + 16m_p)}\left(\frac{2E_e}{m_e}\right)^{1/2} \tag{10}$$

Combining equations (9) and (10) combine to give the maximum recoil energy of an oxygen PKA, which is given numerically by:

$$E_R = 0.68\,\left(\tfrac{E_e}{5keV}\right)^{1/2} eV \tag{11}$$

For a 5 keV electron, this maximum recoil energy is approximately given by 0.68 eV.

Using the semi-empirical power-law scaling of defect creation previously reported [20], given by:

$$N_d = AE_R^{\,m} \tag{12}$$

we calculated the maximum number of (predicted) PKA-defects and plot these in Fig. S3 below. The maximum value for $E_R$ assumed that a 5 keV electron collides head-on with a target nucleus and the parameters A=[ 0.539 1.225 2.104 2.820] and m=[ 0.72 0.80 0.89 0.88] correspond to Si$^{III}$ (oxygen vacancy), O$^{I}$, Si$^{V}$, and O$^{III}$ type defects.

To calculate the fraction of total target volume affected by defects ($f_D$), we assumed that the thickness of the 5 keV electron beam damage was ~250 nanometers, consistent with CASINO calculations (See Fig. S4 and next section) of electron energy loss in SiO$_2$.

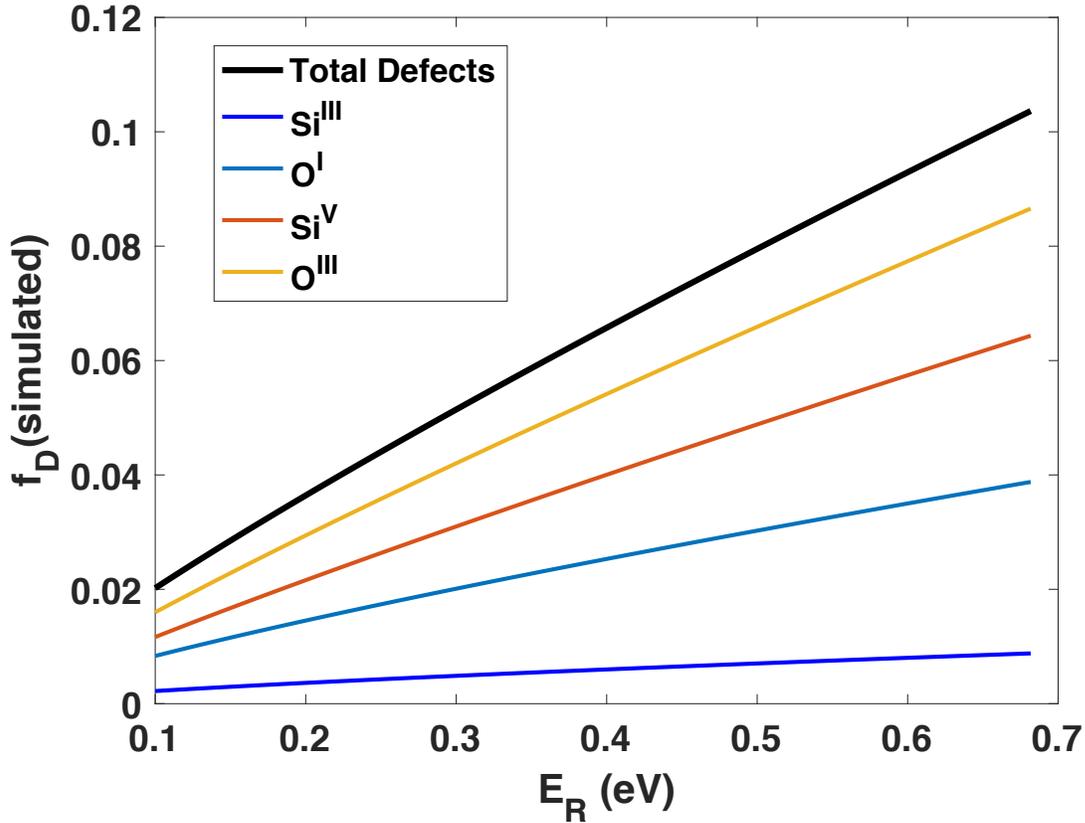

**Fig. S3.** Maximum defect fraction (**f$_D$**) as estimated using the empirical relationship between simulated number of defects per PKA atom and its energy E$_R$ ( $N_d = A E_R{}^m$ ) where $A$ and $m$ are parameters derived from MD simulations of defect formation as reported by [20].

There appears to be semi-quantitative correspondence between the fraction **f$_D$** of the solid that we deduce has been altered by the electron irradiation and exposure to H$_2$O and the total number of defects predicted by MD simulations of radiation damage in quartz. Reconciling these two numbers is beyond the scope of this paper, but could be a fruitful avenue of research as MD simulations and models would benefit from the empirical calibrations made possible by our experimental approach.

### CASINO Simulations of electron beam damage

We estimated the regions of e-beam damage using a Monte-**CA**rlo **SI**mulations of electro**N** trajectory in s**O**lids (CASINO) software v.3.3.0.4 [35]. Figure S4 shows that 5 keV electrons are expected to cause electron beam damage down to a depth of 200-250 nm in SiO$_2$.

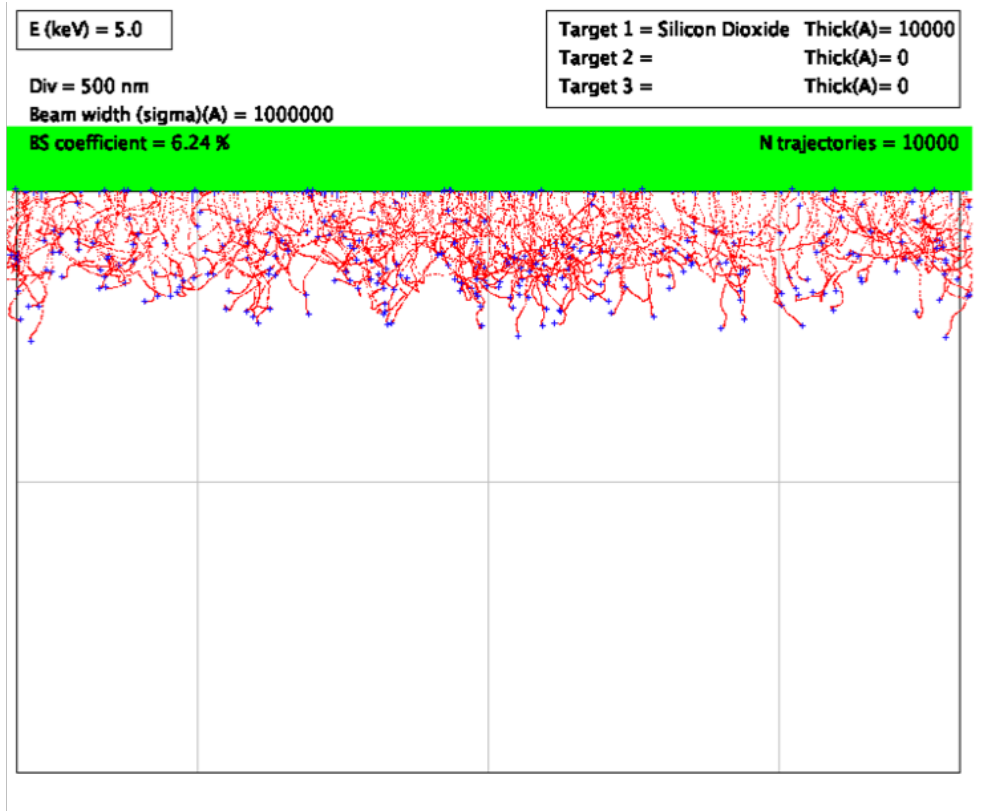

**Fig. S4.** Results of a single scattering simulation of 5 keV electron propagation in $SiO_2$. We note that region heavily influenced by primary electrons is approximately ~200 nm, with blue dots indicating primary electron final location.

**Cosmic-ray Fluxes and Timescale for Defect Formation**

We estimated the timescales for oxygen isotope exchange in the interstellar medium (ISM) and in the early solar system (ESS) by focusing on the production of defects contributed by the proton ($H^+$) components of cosmic-rays (CRs) in each of these environments. For the ISM, we used the cosmic-ray energy spectrum ($S(E_{CR})$) derived from the analysis of the Voyager spacecraft [21] for energies ranging from 1 MeV ($E_{min}$) to 10 GeV ($E_{max}$). While this energy range is rather large, the power-law nature of this distribution translates in a cosmic-ray exposure rate ($\Phi_p = 2.94 \times 10^5$ cm$^{-2}$ s$^{-1}$) in the ISM with an average energy of ~40 MeV.

To estimate the rate of defect production, we first determined how much energy a single cosmic ray with energy $E_{CR}$ loses while traversing a dust grain of radius $r_g$. Using empirical determinations of energy-loss ($\frac{dE(E_{CR})}{dz}$) of protons in $SiO_2$, we estimated the energy deposited in a dust grain of radius $r_g$ as:

$$E_D(E_{CR}, r_g) \simeq 2r_g \frac{dE(E_{CR})}{dz} \qquad (0.13)$$

For example, the energy loss per unit length of a 40 MeV proton in SiO$_2$ is approximately 39.23 MeV/cm which translates into a total energy loss (deposited) of ~790 eV into a 200 nm in radius dust grain. ***It's important to note that cosmic-rays are able to pass-through typical interstellar dust grains without losing a significant amount of their energy (790 eV out of 40 MeV in this case) and the radiation damage caused by cosmic-rays is due to electrons with similar energy to those used in our experiments.***

To estimate the rate of defect formation ($\frac{dN_D}{dt}$), we took two approaches. In the simplest approach, we assume that the production of a single atomic defect takes a threshold energy (or work function) of ($E_{th}$) and the number of defects produced is given by

$$N_D \sim \frac{E_D}{E_{th}} \qquad (0.14)$$

Thus, a single 40 MeV CR proton may be expected to produce ~8 defects in the entire volume of a 200 nm dust grain if E$_{th}$=100 eV. In the second approach, we used the scaling laws of the MD simulations to determine the defect formation rate as a function of the energy of deposition E$_D$. For comparison, this semi-empirical model predicts a total number of ~21 atoms, with ~12 of these being oxygen atoms. **In general, the simple-minded and model approaches produced defect formation rates that were within a factor of 2-3 over the entire CR energy spectrum.**

The overall rate of defect formation was calculated by integrating over the cosmic-ray energy spectrum. This flux weighted defect production rate for a dust grain of radius r$_g$ is given by

$$\frac{dN_D}{dt} = 4\pi \int_{E_{min}}^{E_{max}} N_D(E, r_g) \frac{dJ(E_{CR})}{dE} r_g^2 \, dE_{CR} \qquad (0.15)$$

Finally, the oxygen isotope exchange timescale due to cosmic-ray irradiation is given by:

$$\tau_D(r_g) = \frac{N_O(r_g)}{\dfrac{dN_D(r_g)}{dt}} \qquad (0.16)$$

Where $N_O$ is the number of oxygen atoms found within the dust grain. For simplicity, we assumed that the number of oxygen atoms in a dust grain of radius r$_g$ is given by:

$$N_O(r_g) \sim 2 \times \frac{\frac{4\pi}{3} \rho_g r_g^3}{60 m_h} \qquad (0.17)$$

for a pure SiO$_2$ dust grain composition where each SiO$_2$ molecular unit has a single Si atom and two [16]O atoms.

A summary of the major variables and terms used in these calculations is given in Table S2 below.

**Cosmic Ray Energy Spectra Used**

To estimate the isotope exchange timescales in the ISM and early solar system, contemporary solar system and early solar system we used the cosmic-ray electron spectrum $\frac{dJ}{dE}$ as described in [22] and the corresponding energy loss functions described by [36]   This energy spectrum is reproduced in Figure S5.

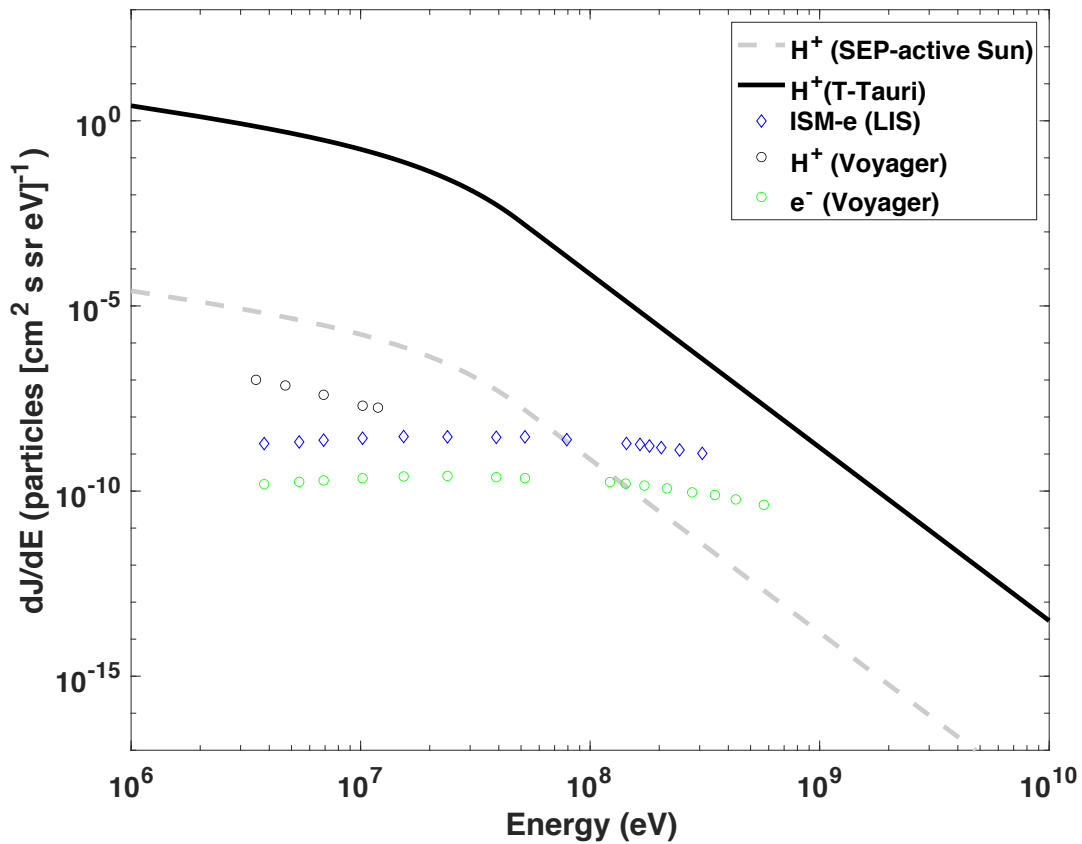

**Fig. S5.** Summary of cosmic-ray intensity (J) spectra used for estimates of CR fluences, energy deposition, and defect formation timescales. Proton and electron cosmic-ray spectral intensities as reported by Voyager [21]. SEP H[+] events reported by [22] (SEP-active Sun) were scaled by a factor of $10^5$ to estimate the T-Tauri H+ spectral intensities as described by [25] . For comparison, the electron component of galactic cosmic rays (ISM-e) was reproduced as described in [37].

A separate source of energetic particles with energies > 10 MeV are Solar Energetic Particle events (SEPs) associated with Coronal Mass Ejections. The abundance of short-lived radionuclides in meteorites cannot be explained by contemporary solar CR fluxes ($\Phi_p \sim 150$ protons cm$^{-2}$ s$^{-1}$). However, similar to T-Tauri stars characterized by their enhanced solar activities, it is likely that the young Sun in the early solar system may have been more active by up to a factor of $\sim 10^5$. To estimate the solar CR flux experienced by dust grains in the early solar system, we followed an approach similar to [25] and scaled the cycle averaged CR energy spectrum for event 1 in [22] by a factor of $\sim 10^5$ to give us a T-Tauri solar cosmic ray energy spectrum for protons ($S_{TT}(E_{CR})$) and a total flux of protons with energies > 10 MeV of $\Phi_p(E > 10 MeV) = 1.5x10^7$ cm$^{-2}$ s$^{-1}$ at 1 a distance of 1 AU from the Sun. The CR spectrum (T-Tauri) used in our calculations of the oxygen isotope exchange timescales is reproduced in **Figure S5** while the timescales as a function of dust grain size and distance from the Sun are plotted in the main text **(Figure 5)**.

**Estimating Defect and Oxygen Exchange Rates and Timescales**

Table S2  Summary of major variables and quantities used to calculate defect formation rates in dust grain volumes

| Variable | Description | Description | Units |
|---|---|---|---|
| $S(E_{CR})$ | $= \dfrac{dJ}{dE}$ | Cosmic-ray Energy Spectrum | Particles /[MeV cm$^2$ sr s] |
| $\Phi_p$ | $= = 4\pi \int_{E_{min}}^{E_{max}} S(E_{CR}) dE_{CR}$ | Total Number of Particles between E$_{min}$ and E$_{max}$ | Particles /(cm$^2$ s) |
| $\dfrac{dE(E_{CR})}{dz}$ | Empirical look up table (Ref. #) | Energy-loss per unit length of particle (H$^+$,e$^-$) | eV/cm |
| $E_D(E_{CR}, r_g)$ | $\simeq 2 r_g \dfrac{dE(E_{CR})}{dz}$ | Amount of energy deposited in a dust grain of radius r$_g$ | eV |
| $N_D(E_{CR}, r_g)$ | $= F(E_D(E_{CR}, r_g))$ | Number of defects created in a dust grains with radius r$_g$ per cosmic ray particle with energy E$_{CR}$ | Number |
| $\dfrac{dN_D(r_g, E_{min}, E_{max})}{dt}$ | $= 4\pi \int_{E_{min}}^{E_{max}} N_D(E, r_g) \dfrac{dJ(E_{CR})}{dE} r_g^2 dE_{CR}$ | Rate of defect formation in dust grain of radius r$_g$ | Number/s |

| $\tau_D$ | $= \dfrac{N_O(r_g)}{\dfrac{dN_D(r_g)}{dt}}$ | Oxygen isotope exchange timescale | s |
| --- | --- | --- | --- |

We used the Voyager results of [21] to calculate cosmic-ray proton (H$^+$) and electron (e$^-$) fluxes and the timescale ($\tau_D$) needed for cosmic rays to break a significant number of bonds in interstellar dust grains. Using data of galactic cosmic ray electrons and proton intensities (J) as a function of energy E presented in [21] we calculated the number of defects produced by CR protons and electrons as follows. The amount of energy deposited per CR ion into a dust grain of diameter $D_g (= 2r_g)$ is given by:

$$\Delta E_{CR} = \left( \frac{dE}{dx}(E_{CR}) \right) D_g \qquad (18)$$

where $\dfrac{dE}{dx}$ is the energy-loss or a cosmic-ray ion per unit length traversed in the dust grain material.

Every discrete CR ion interaction is expected to lead to energetic electron production within the dust grain volume. We estimate the number of defects created by these interactions by applying equation (12) to estimate the number of defects as:

$$N_D(\Delta E_{CR}) = A(\Delta E_{CR})^m \qquad (19)$$

Therefore, the rate of defect formation in a dust grain of radius $r_g$ caused by CRs of energy E is given by:

$$\left( \frac{dN_D}{dEdt} \right)_{Total} = J(E) \cdot N_D(\Delta E_{CR}) \cdot 4\pi r_g^2 \qquad (20)$$

where $J(E)$ is given in units of number of cosmic rays per steradian per MeV per second.

We integrated this expression to obtain the total defect production rate for all cosmic-ray energies in a dust grain of radius r$_g$ as follows:

$$\left( \frac{dN_D}{dt} \right)_{Total} = \sum_{E_{CR}} J(E_{bin}) \cdot N_D(\Delta E_{CR}) \cdot 4\pi r_g^2 \cdot \delta E \qquad (21)$$

where $\delta E = (E_{min} - E_{max})$, $E_{bin} = E_{min} + (E_{max} - E_{min})/2$, and the sum is performed over the range of energies for a given cosmic-ray energy intensity spectrum $J(E)$ as shown in Figure S5.

Finally, the characteristic timescale (1/e) for significant defect formation in a dust grain of radius r$_g$ is given by:

$$\tau_D = \frac{N_{O-atoms}}{\left(\dfrac{dN_D}{dt}\right)_{Total}} \tag{22}$$

where $N_{O-atoms}$ is the number of oxygen atoms in an $SiO_2$ dominated dust grain.

For the ISM, we tabulated the rates of defect formation as well as the corresponding timescales separately for protons and electrons and combined these to obtain the total defect timescale as:

$$\tau_D(total) = \left(\frac{1}{\tau_D(H^+)} + \frac{1}{\tau_D(e^-)}\right)^{-1} \tag{23}$$

The results of these numerical calculations of defect formation timescales for dust grains as a function of their radius ($r_g$) are given in Figure 5 in the main text.

**Oxygen Isotopic Evolution and Steady State Solution**

We consider a water ice ($N_{ice}$) and a solid ($N_s$) reservoirs that can exchange oxygen atoms and seek solutions to their steady-state isotopic compositions. Here we focus on the exchange of the anomalous component of oxygen content of each as measured by $\Delta^{17}O$. To do this, we consider isotopic fluxes ($= \frac{dN_i}{dt}\Delta^{17}O_i$) into and out of each reservoir. The time evolution of the isotopic composition of the $H_2O$ ice reservoir is given by:

$$\frac{d}{dt}\left(\Delta^{17}O_{ice}(t)N_{ice}\right) = \frac{dN}{dt}\Delta^{17}O_s(t) - \frac{dN}{dt}\left(\Delta^{17}O_{ice}(t) + \Delta^{17}O_{max}\right) \tag{24}$$

where the isotopic composition of atoms moving from the solid to the ice phase is equal to the instantaneous isotopic composition of the solid ($\Delta^{17}O_s$) and the isotopic composition of atoms flowing from the ice back into the solid reservoir is given by the isotopic composition of the ice ($\Delta^{17}O_{ice}(t)$) *plus a constant* $\Delta^{17}O_{max}$. After enough time has elapsed compared to the defect formation time ($t \gg t_D$), the isotopic composition of the ice and solid are expected to reach steady-state and the left-hand size of the above equation is equal to 0, giving us the following:

$$0 = \frac{dN}{dt}\Delta^{17}O_s(t_{ss}) - \frac{dN}{dt}\left(\Delta^{17}O_{ice}(t_{ss}) + \Delta^{17}O_{max}\right) \tag{25}$$

where $t_{ss}$ denotes steady state isotopic compositions of each reservoir. It is easy to show that the above equation implies that the isotopic difference between the solid and water ice reservoir is given by:

$$\Delta^{17}O_s(t_{ss}) - \Delta^{17}O_{ice}(t_{ss}) = \Delta^{17}O_{max} \tag{26}$$

For example, we note that if there is no anomalous isotope exchange ($\Delta^{17}O_{max}$=0), $\Delta^{17}O_s$ = $\Delta^{17}O_{ice}$ as expected. Figure 5 in main text shows the results of numerical simulations of the isotopic composition of the dust reservoir vs. time for different isotope exchange rates.

## References and Notes:

1    Williams, J. P. & Cieza, L. A. Protoplanetary Disks and Their Evolution. *Annual Review of Astronomy and Astrophysics, vol. 49, issue 1, pp. 67-117* **49**, 67-117, doi:10.1146/annurev-astro-081710-102548 (2011).

2    McKeegan, K. D. *et al.* The Oxygen Isotopic Composition of the Sun Inferred from Captured Solar Wind. *Science* **332**, 1528, doi:10.1126/science.1204636 (2011).

3    Jones, A. P., Tielens, A. G. G. M., Hollenbach, D. J. & McKee, C. F. Grain Destruction in Shocks in the Interstellar Medium. *Astrophysical Journal v.433, p.797* **433**, 797, doi:10.1086/174689 (1994).

4    Nittler, L. R. & Ciesla, F. Astrophysics with Extraterrestrial Materials. *Annual Review of Astronomy and Astrophysics* **54**, 53-93, doi:10.1146/annurev-astro-082214-122505 (2016).

5    Thiemens, M. H., Chakraborty, S. & Dominguez, G. The Physical Chemistry of Mass-Independent Isotope Effects and Their Observation in Nature. *Annu. Rev. Phys. Chem.* **63**, 155-177, doi:papers2://publication/doi/10.1146/annurev-physchem-032511-143657 (2012).

6    Franchi, I. A., Wright, I. P., Sexton, A. S. & Pillinger, C. T. The oxygen-isotopic composition of Earth and Mars. *Meteoritics & Planetary Science* **34**, 657-661, doi:10.1111/j.1945-5100.1999.tb01371.x (1999).

7    McKeegan, K. D. *et al.* The Oxygen Isotopic Composition of the Sun Inferred from Captured Solar Wind. *Science* **332**, 1528-1532, doi:10.1126/science.1204636 (2011).

8    Yurimoto, H. *et al.*   849.

9    Gounelle, M., Krot, A. N., Nagashima, K. & Kearsley, A. EXTREME16O ENRICHMENT IN CALCIUM-ALUMINUM-RICH INCLUSIONS FROM THE ISHEYEVO (CH/CB) CHONDRITE. *The Astrophysical Journal* **698**, L18-L22, doi:10.1088/0004-637x/698/1/l18 (2009).

10   Lyons, J. R. & Young, E. D. CO self-shielding as the origin of oxygen isotope anomalies in the early solar nebula. *Nature* **435**, 317-320 (2005).

11   Yurimoto, H. & Kuramoto, K. Molecular Cloud Origin for the Oxygen Isotope Heterogeneity in the Solar System. *Science* **305**, 1763-1766, doi:10.1126/science.1100989 (2004).

12   Lyons, J. *et al. Timescales for the evolution of oxygen isotope compositions in the solar nebula*. Vol. 73 (2009).

13   Thiemens, M. H. & Heidenreich, J. E., III. The mass-independent fractionation of oxygen - A novel isotope effect and its possible cosmochemical implications. *Science* **219**, 1073-1075 (1983).

14   Chakraborty, S., Yanchulova, P. & Thiemens, M. H. Mass-Independent Oxygen Isotopic Partitioning During Gas-Phase SiO2 Formation. *Science* **342**, 463, doi:10.1126/science.1242237 (2013).

15    . (!!! INVALID CITATION !!! 16,17).

16    Abplanalp, M. J. *et al.* A study of interstellar aldehydes and enols as tracers of a cosmic ray-driven nonequilibrium synthesis of complex organic molecules. *Proceedings of the National Academy of Sciences* **113**, 7727, doi:10.1073/pnas.1604426113 (2016).

17    Dominguez, G. A Heterogeneous Chemical Origin for the $^{16}O$-enriched and $^{16}O$-depleted Reservoirs of the Early Solar System. *The Astrophysical Journal Letters* **713**, L59-L63 (2010).

18    Berger, M. J., Coursey, J.S., Zucker, M.A., and Chang, J. . *STAR, PSTAR, and ASTAR: Computer Programs for Calculating Stopping-Power and Range Tables for Electrons, Protons, and Helium Ions (version 1.2.3)*, <http://physics.nist.gov/Star> (2005).

19    Wood, S. *et al.* Simulation of Radiation Damage in Solids. *IEEE Transactions on Nuclear Science* **28**, 4107-4112, doi:10.1109/TNS.1981.4335684 (1981).

20    Wang, B., Yu, Y., Pignatelli, I., Sant, G. & Bauchy, M. Nature of radiation-induced defects in quartz. *The Journal of Chemical Physics* **143**, 024505, doi:10.1063/1.4926527 (2015).

21    Cummings, A. C. *et al.* GALACTIC COSMIC RAYS IN THE LOCAL INTERSTELLAR MEDIUM:VOYAGER 1OBSERVATIONS AND MODEL RESULTS. *The Astrophysical Journal* **831**, 18, doi:10.3847/0004-637x/831/1/18 (2016).

22    Mewaldt, R. A. *et al.* Proton, helium, and electron spectra during the large solar particle events of October–November 2003. *Journal of Geophysical Research: Space Physics* **110**, doi:https://doi.org/10.1029/2005JA011038 (2005).

23    Chaussidon, M. & Gounelle, M. in *Meteorites and the Early Solar System II*    323 (2006).

24    Sossi, P. A. *et al.* Early Solar System irradiation quantified by linked vanadium and beryllium isotope variations in meteorites. *Nature Astronomy* **1**, 0055, doi:10.1038/s41550-017-0055 (2017).

25    Rab, C. *et al.* Stellar energetic particle ionization in protoplanetary disks around T Tauri stars. *A&A* **603** (2017).

26    Feigelson, E. D., Garmire, G. P. & Pravdo, S. H. Magnetic Flaring in the Pre-Main-Sequence Sun and Implications for the Early Solar System. *The Astrophysical Journal* **572**, 335-349, doi:10.1086/340340 (2002).

27    Asplund, M., Grevesse, N., Sauval, A. J. & Scott, P. The Chemical Composition of the Sun. *Annual Review of Astronomy and Astrophysics* **47**, 481-522, doi:10.1146/annurev.astro.46.060407.145222 (2009).

28    Min, M. *et al.* The shape and composition of interstellar silicate grains. *Astronomy and Astrophysics* **462**, 667-676, doi:10.1051/0004-6361:20065436 (2007).

29    Franchi, I. A. Oxygen Isotopes in Asteroidal Materials. *Reviews in Mineralogy and Geochemistry* **68**, 345-397, doi:10.2138/rmg.2008.68.13 (2008).

30    Krot, A. N. Refractory inclusions in carbonaceous chondrites: Records of early solar system processes. *Meteoritics & Planetary Science* **54**, 1647-1691, doi:https://doi.org/10.1111/maps.13350 (2019).

31    Marcus, R. A. Mass-independent isotope effect in the earliest processed solids in the solar system: A possible chemical mechanism. *The Journal of Chemical Physics* **121**, 8201-8211, doi:10.1063/1.1803507 (2004).

32    Smith, R. L., Pontoppidan, K. M., Young, E. D., Morris, M. R. & van Dishoeck, E. F. High-Precision C$^{17}$O, C$^{18}$O, and C$^{16}$O Measurements in Young Stellar Objects:

Analogues for CO Self-shielding in the Early Solar System. *Astrophysical Journal* **701**, 163-175 (2009).

33    Simon, J. I. *et al.* Oxygen Isotope Variations at the Margin of a CAI Records Circulation Within the Solar Nebula. *Science* **331**, 1175, doi:10.1126/science.1197970 (2011).

34    Li, S. *et al.* Direct evidence of surface exposed water ice in the lunar polar regions. *Proceedings of the National Academy of Sciences* **115**, 8907, doi:10.1073/pnas.1802345115 (2018).

35    Drouin, D. *et al.* CASINO V2.42—A Fast and Easy-to-use Modeling Tool for Scanning Electron Microscopy and Microanalysis Users. *Scanning* **29**, 92-101, doi:https://doi.org/10.1002/sca.20000 (2007).

36    4. Charged-Particle Stopping Powers and Related Quantities. *Journal of the International Commission on Radiation Units and Measurements* **14**, 21-30, doi:10.1093/jicru/ndw031 (2016).

37    Potgieter, M. S., Vos, E. E., Munini, R., Boezio, M. & Felice, V. D. MODULATION OF GALACTIC ELECTRONS IN THE HELIOSPHERE DURING THE UNUSUAL SOLAR MINIMUM OF 2006–2009: A MODELING APPROACH. *The Astrophysical Journal* **810**, 141, doi:10.1088/0004-637x/810/2/141 (2015).

**Acknowledgments**

Ezra Benitez, Charisa Boyer, Elizabeth Christiansen, Manesseh Park, Jonathan Naveh, Carina Maciel are thanked for helping to construct, characterize, and test the Isotopic Characterization Experimental (ICE) Apparatus at CSUSM. A. Westphal is acknowledged for his helpful discussions of cosmic-ray abundances.

**Author contributions:** Writing**,** Conceptualization, Funding Acquisition of ICE experiments, and Methodology was done by G.D. Investigations at CSUSM were supervised by G.D. and were conducted by G.D., L.T., and J.L. Data Curation and Formal Analysis of SIMS data was done by M.C.L and G.D. Funding Acquisition of SIMS facilities was done by K.M.

**Funding:** G.D acknowledges support from NSF's Astronomy and Astrophysics (1616992) and NASA's Cosmochemistry (NNX13AN25G) programs. Financial support for L.Tafla and J. Lucas was provided by the San Diego Foundation and NSF. The ion microprobe facility at UCLA is partially supported by the Instrumentation and Facilities Program, Division of Earth Sciences, NSF (EAR-1339051 and EAR- 1734856).

**Competing interests:** Authors declare no competing interests

**Data and materials availability:** All data is available in the main text or the supplementary materials.

**Corresponding author:** Gerardo Dominguez (gdominguez@csusm.edu)

**Additional Information:**

Supplementary Information is available for this paper.

Correspondence and requests for materials should be addressed to G. Dominguez.

Reprints and permissions information is available at www.nature.com/reprints.

[i] A note on $\delta^x O$ notation: It is typical in isotope geochemistry to denote the abundance of an isotope of an element using "delta" notation. For any element **X** with major isotope of mass i ($^i\mathbf{X}$) and minor isotope of mass j ($^j\mathbf{X}$), comparison of the relative ratio of these two isotopes compared to a well-known standard is quantified as a deviation from this standard as:

$$\delta^i X = [(Rsample(i,j)/Rstandard(i,j) -1 ]x1000$$

where the ratio of abundances of isotopes is given by $R_y(i,j)=[^j X]/[^i X]$ (y=sample, standard).

In the case of oxygen, the anomalous or mass-independent $^{17}O$ content of an object is quantified as the deviation from the expected mass-dependent behavior as:

$$\Delta^{17}O = \delta^{17}O(meas.) - 0.52 \text{ x } \delta^{18}O(meas.)$$